\documentclass{ws-ijmpcs}
\begin{document}

\newcommand{\okgr}{\Omega_{\,\rm K}}

\markboth{R. Mellinger {\it et al.}}
         {Rotation-driven phase transitions in the cores of pulsars}

%
\catchline{}{}{}{}{}
%

\title{Rotation-driven phase transitions in the cores of pulsars}

\author{Richard D. Mellinger Jr$^{a,\dag}$, Fridolin
  Weber$^{a,b,\ddag}$, William Spinella$^{a,*}$,
  G.A. Contrera$^{c,d,e,\S}$, and M. Orsaria$^{a,c,d,\P}$
\vspace{6px}}

\address{$^a$Department of Physics, San Diego State University, 5500
  Campanile Drive, \\ San Diego, CA 92182, USA.\\
$^b$Center for Astrophysics and Space Sciences, University of
  California San Diego, \\ La Jolla, CA 92093, USA.\\
$^c$National Scientific and Technical Research Council (CONICET),
  Godoy Cruz 2290, \\ 1425 Buenos Aires, Argentina. \\
$^d$Grupo de Gravitaci\'{o}n, Astrof\'{\i}sica y Cosmolog\'{\i}a,
  Facultad de Ciencias Astron\'{o}micas \\ y Geof\'{\i}sicas, Universidad
  Nacional de La Plata, La Plata, Argentina. \\
$^e$Instituto de F\'{\i}sica La Plata, National Scientific and
  Technical Research Council (CONICET), Universidad Nacional de La
  Plata, La Plata , Argentina. \\
$^{\dag}$imasillypirate@gmail.com, $^{\ddag}$fweber@mail.sdsu.edu,
$^*$spinella@rohan.sdsu.edu
$^{\S}$contrera@fisica.unlp.edu.ar,
$^{\P}$morsaria@fcaglp.unlp.edu.ar
}

\maketitle

\begin{history}
\received{Day Month Year}
\revised{Day Month Year}
\published{Day Month Year}
\end{history}

\begin{abstract}
  In this paper, we discuss the impact of rotation on the particle
  composition of rotating neutron stars (pulsars). Particular emphasis
  is put on the formation of quark matter during stellar spin-down,
  driven by continuous gravitational compression. Our study is based on
  modern models for the nuclear equation of state whose parameters are
  tightly constrained by nuclear data, neutron star masses, and the
  latest estimates of neutron star radii.

\end{abstract}

\ccode{PACS numbers:25.75.Nq, 26.60.-c, 26.60.Kp, 97.60.Gb}

\section{Introduction}

Most of the neutron star calculations reported in the literature have
been performed for non-rotating, spherically symmetric stellar
configurations, whose properties are uniquely determined by the
Oppenheimer-Volkoff equation.\cite{glen97:book,weber99:book} However a
steady increase in the number of observed rotating neutron stars with
rotational periods in the millisecond range has renewed considerable
interest in the influence and implications of rapid rotation on the
properties of neutron stars.  Details about these equations of state
(EoS) can be found in the recent paper by Mellinger {\it et
  al.\,}\cite{mellinger17:a} and in the contribution of Contrera {\it et
  al.\,}\cite{contrera17:IWARA} contained elsewhere in this volume.

\section{Rotating Neutron Stars in the Framework of General Relativity}\label{sec:GR}

The fact that rotation deforms neutron stars, stabilizes them against
collapse, and drags along the local inertial frames inside and outside
of them so that they co-rotate with the stars, renders the
construction of models of rotating neutron stars rather complicated. A
suitable ansatz for the line element (that is, the components of the
metric tensor) has the form\cite{weber99:book}
\begin{eqnarray}
  d s^2 = - e^{2\nu} dt^2 + e^{2\psi} (d\phi - \omega dt)^2 + e^{2\mu}
  d\theta^2 + e^{2\lambda} dr^2 \, ,
\label{eq:metric} 
\end{eqnarray} where   $\nu$, $\psi$,  $\mu$ and
$\lambda$ denote metric functions and $\omega$ is the angular velocity
of the local inertial frames. All these quantities depend on the
radial coordinate $r$, the polar angle $\theta$, and implicitly on the
star's angular velocity $\Omega$.  The metric functions and the frame
dragging frequencies are to be computed from Einstein's field
equation,
\begin{equation} 
R^{\kappa\sigma} - \frac{1}{2} R g^{\kappa\sigma} = 8 \pi
T^{\kappa\sigma} \, ,
\label{eq:einstein}
\end{equation}
where $T^{\kappa\sigma} = T^{\kappa\sigma}(\epsilon, P(\epsilon))$
denotes the energy momentum tensor of the stellar matter, whose
equation of state is given by $P(\epsilon)$. The other quantities in
Eq.\ (\ref{eq:einstein}) are the Ricci tensor $R^{\kappa\sigma}$, the
curvature scalar $R$, and the metric tensor, $g^{\kappa\sigma}$. No
\begin{figure}[tb]
\begin{center}
  \includegraphics[scale=0.24]{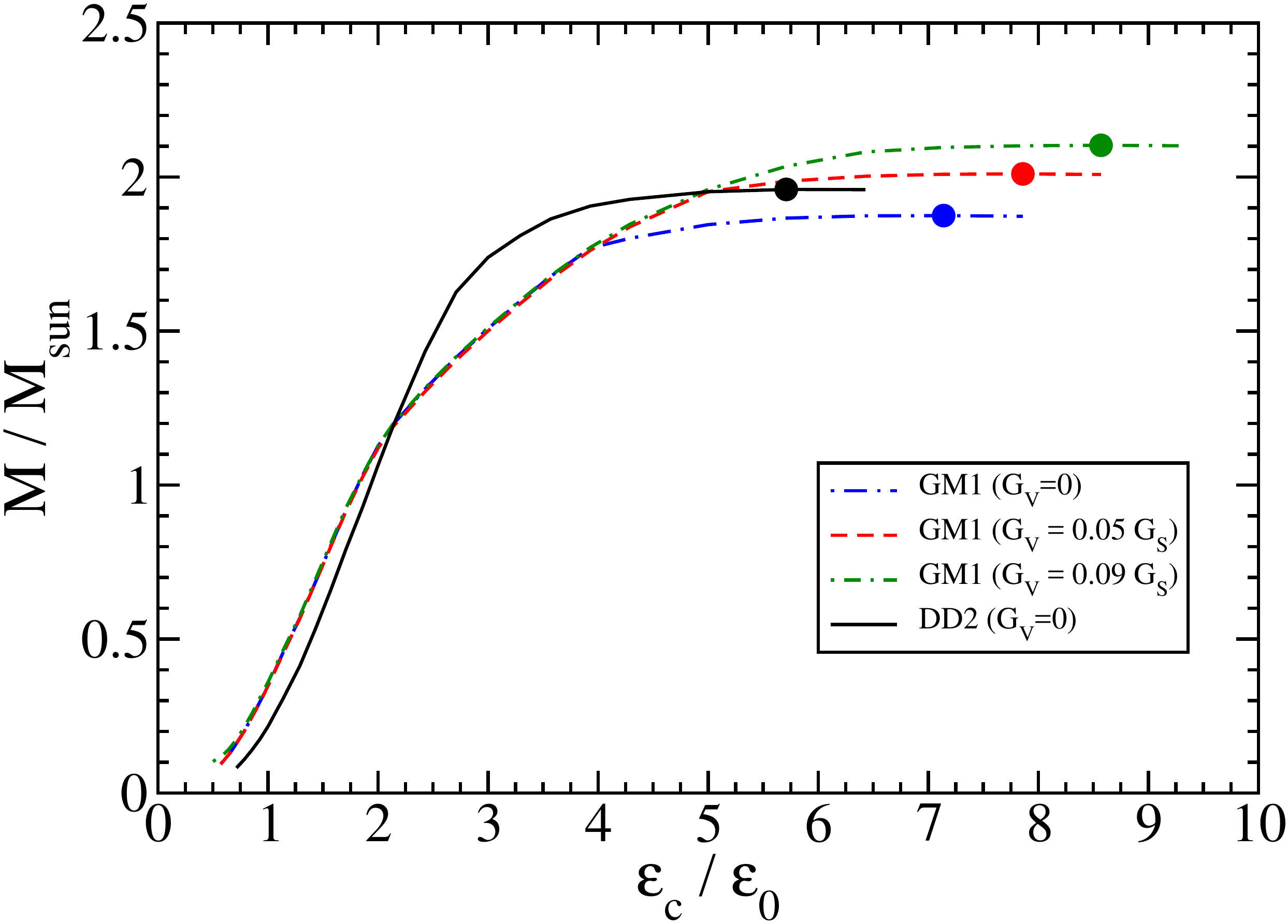}
  \includegraphics[scale=0.24]{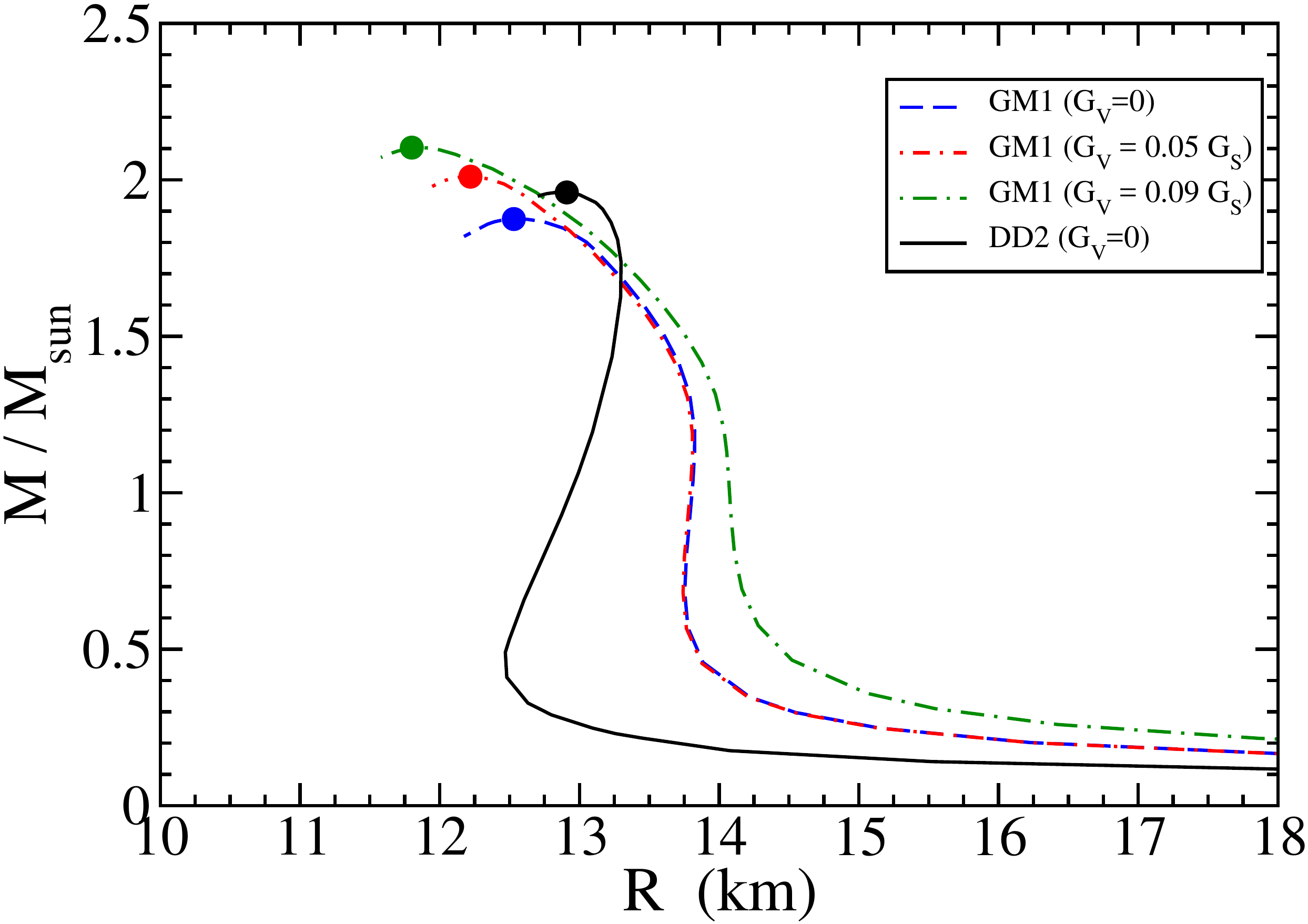}
\end{center}
\caption{Mass--central density (left) and mass--radius relationships
  of non-rotating neutron stars for several nuclear EoS.
  ($\epsilon_0 = 140$ MeV/fm$^3$ denotes the density of infinite
  nuclear matter.) The solid dots mark the maximum-mass star of each
  stellar sequence. (Figure from Ref.~3.)}
  \label{fig:MecMR}
\end{figure}
simple stability criteria are known for rapidly rotating stellar
configurations in general relativity. However, an absolute limit on
rapid rotation is set by the onset of mass shedding from the equator
of a rotating star. The corresponding rotational frequency is known as
the Kepler frequency, ${\Omega_{\,\rm K}}$. In classical mechanics,
the expression for the Kepler frequency, determined by the equality
between the centrifugal force and gravity, is given by ${\Omega_{\,\rm
    K}} = \sqrt{M/R^3}$. Its general relativistic counterpart, which
is obtained from $\delta \int ds^2=0$ evaluated in the star's
equatorial plane, reads\cite{weber99:book,friedman86:a}
\begin{eqnarray}
  {\Omega_{\,\rm K}} = \omega +\frac{\omega_{,r}} {2\psi_{,r}} +
  e^{\nu -\psi} \sqrt{ \frac{\nu_{,r}} {\psi_{,r}} +
    \Bigl(\frac{\omega_{,r}}{2 \psi_{,r}} e^{\psi-\nu}\Bigr)^2 } \, ,
\label{eq:okgr}  
\end{eqnarray}
where ${_{,r}} \equiv \partial/\partial r$. Equation (\ref{eq:okgr}) is
to be evaluated self-consistently at the equator of a rotating neutron
star. The Kepler period follows from Eq.\ (\ref{eq:okgr}) as
${P_{\,\rm K}} = {{2 \pi} / {\Omega_{\,\rm K}}}$. For typical neutron
star matter equations of state, the Kepler period obtained for $1.4\,
M_\odot$ neutron stars is typically around
1~ms.\cite{glen97:book,weber99:book} An exception to this are strange
quark matter stars. Since they are self-bound, they tend to possess
smaller radii than neutron stars, which are bound by gravity
only. Because of their smaller radii, strange stars can withstand mass
shedding from the equator down to rotational periods of around
0.5~ms.\cite{glen92:crust,glen92:limit}

\section{Properties of Rotating Neutron Stars}\label{sec:Results}

A mass increase of up to $\sim 20$\% is typical for rotation at
$\okgr$ (cf.\ Fig.\ \ref{fig:MecMR}). Because of rotation, the
equatorial radii increase by several kilometers, while the polar radii
become smaller by several kilometers. The ratio between both radii is
around 2/3, except for rotation close to the Kepler frequency. The
most rapidly rotating, currently known neutron star is pulsar PSR
J1748-2446ad, which rotates at a period of 1.39~ms,
(719~Hz)\cite{Hessels:2006ze} well below the Kepler frequency for most
neutron star equations of state.\cite{glen97:book,weber99:book}

The density change in the core of a neutron star whose frequency
varies from $0 \leq \Omega \leq {\Omega_{\,\rm K}}$ can be as large as
60\%.\cite{weber99:book,weber05:a} This suggests that rotation may
drive phase transitions and/or cause significant compositional changes of
the matter in the cores of neutron
stars.\cite{glen97:book,weber99:book,weber05:a}

\begin{figure}[tb]
\begin{center}
    \includegraphics[scale=0.30]{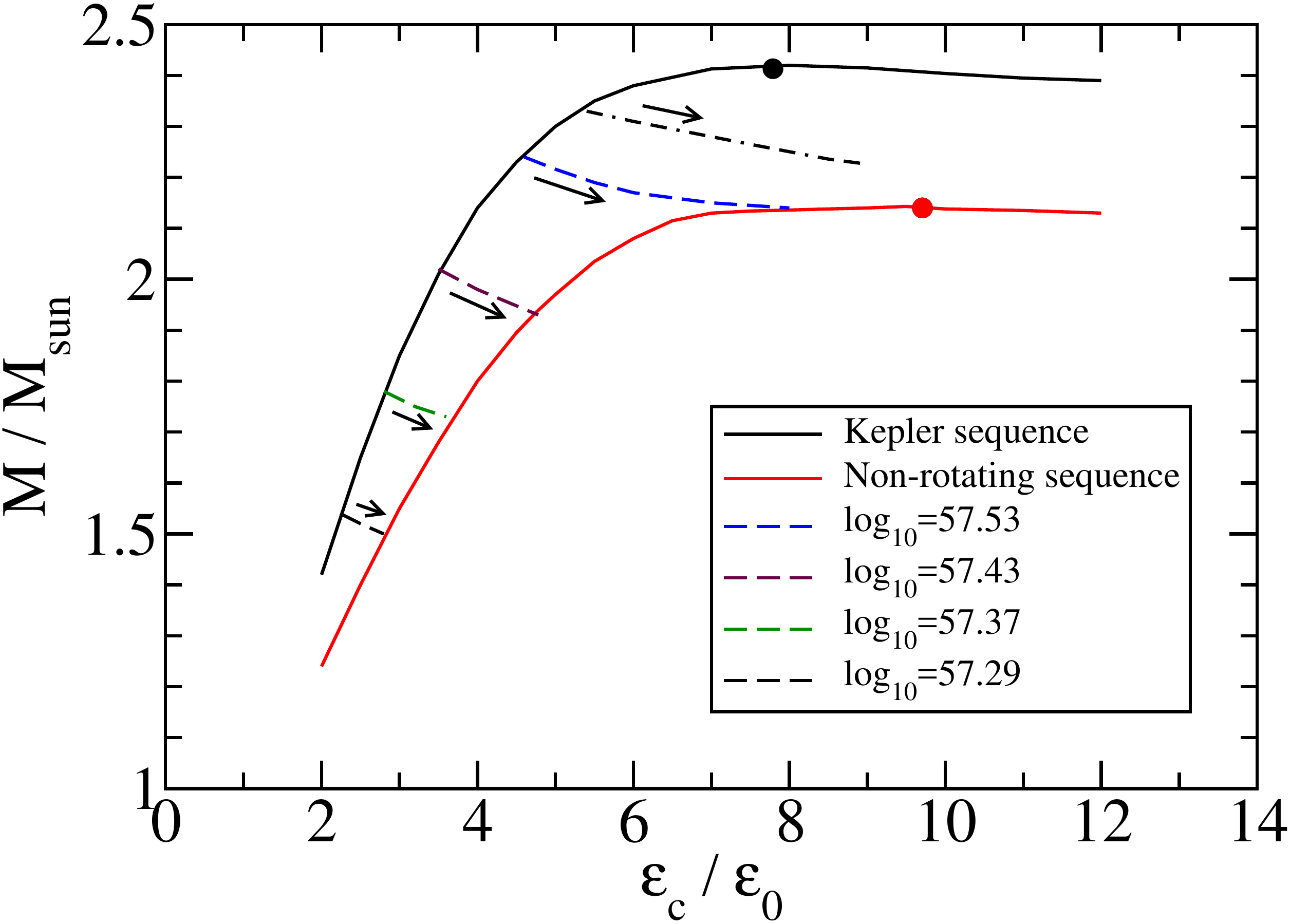}
\end{center}
\caption{Gravitational mass as a function of central stellar density
  of non-rotating and rotating neutron stars for the nuclear EoS
  GM1. Shown are several stellar paths (marked with arrows) that would
  be followed by neutron stars with a constant baryon number, $A$, as
  they spin down from their respective Kepler frequencies (black
  curve), $\Omega_{\rm K}$, to zero frequency (red curve).  (See text
  for more details.)}
  \label{fig:MvsecA}
\end{figure}

Figure \ref{fig:MvsecA} shows the mass versus central density
relationships for both neutron stars spinning at their Kepler
frequencies as well as non-rotating neutron stars. As one moves along
either of these two lines, the baryon number changes, increasing with
larger central density. If one were to assume that a neutron star were
secluded as it spins down (thus, not gaining or losing material), it
would follow a path of
constant baryon number, $A$, from the Kepler frequency curve down to
the non-rotating one. Five such paths are depicted in
Fig.\ \ref{fig:MvsecA}. One sees that the Kepler frequency curve
allows stable solutions with higher baryon numbers than there are
stable solutions for the non-rotating case (dash-dotted curve in
Fig.\ \ref{fig:MvsecA}).  Neutron stars of this type are called
supra-massive rotating neutron stars (SURONS).\cite{mellinger17:a} It
is thought that supra-massive rotating neutron stars collapse into
black holes and have been a proposed source of fast radio
bursts.\cite{falcke14:a}

\begin{figure}[b]
\begin{center}
\includegraphics[scale=0.24]{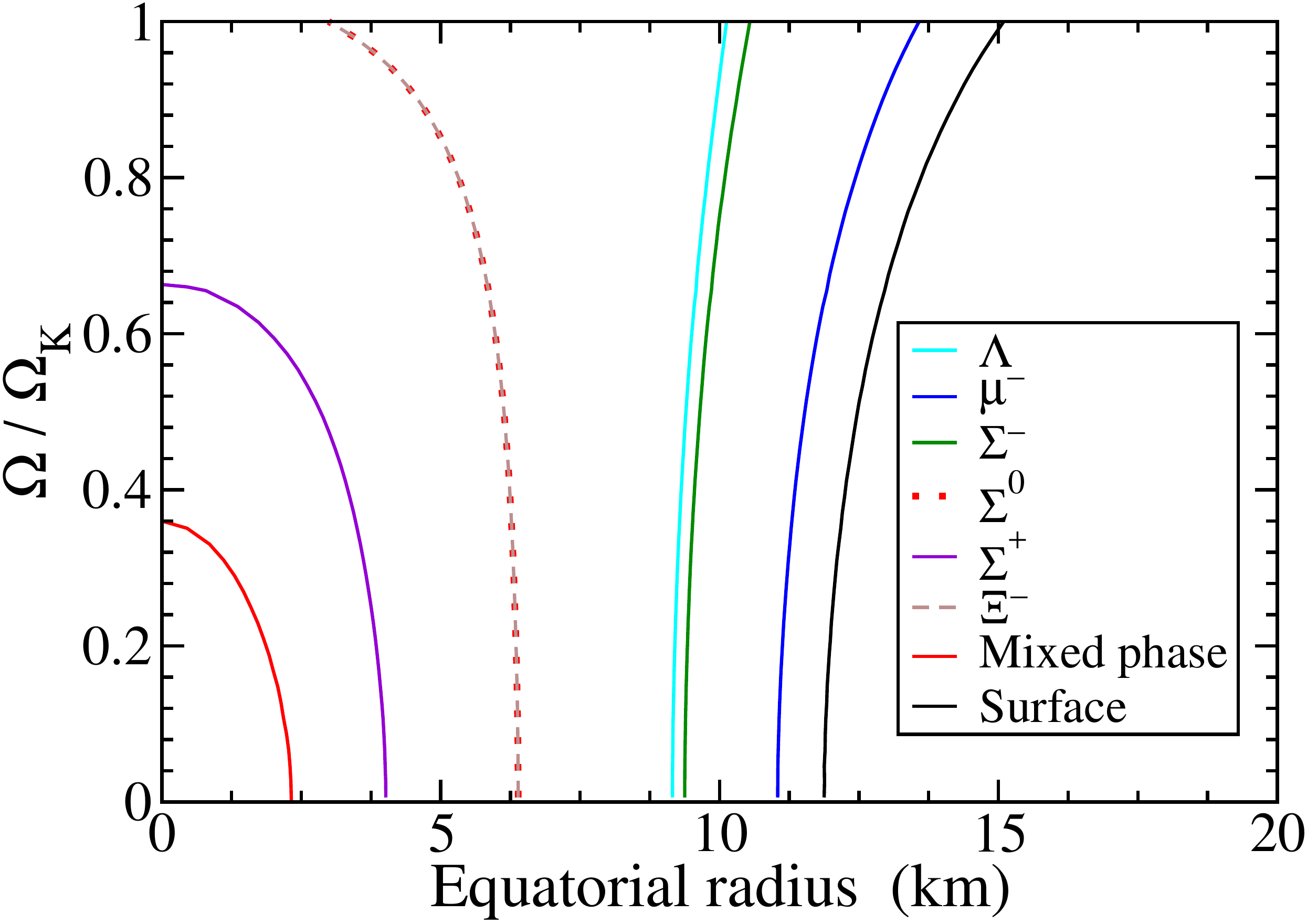}
\includegraphics[scale=0.24]{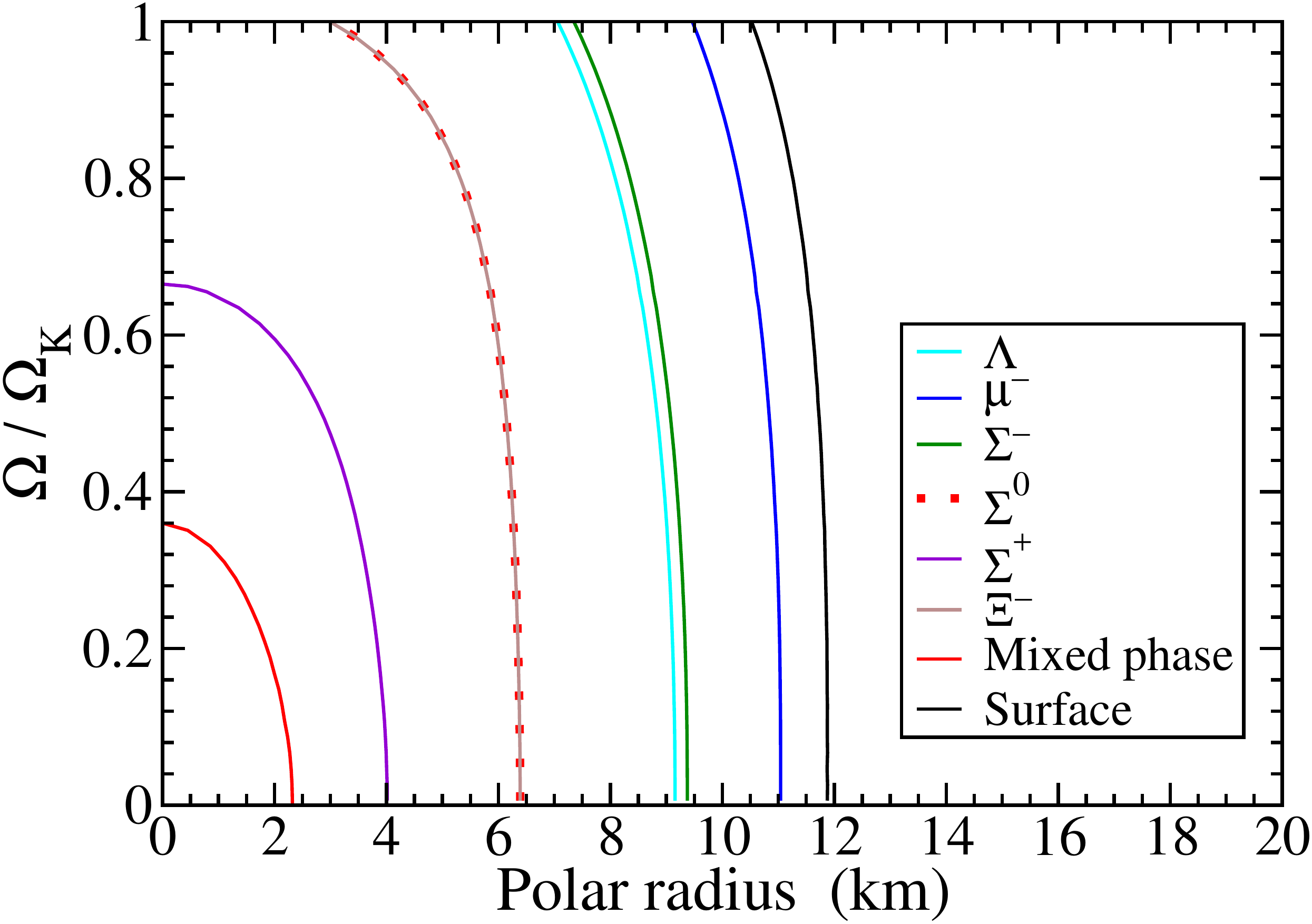}
\caption{Particle populations inside of rotating neutron stars, in
  equatorial (left) and polar (right) directions, computed for the GM1
  EoS. The stellar frequency, $\Omega$, ranges from zero to the
  Kepler frequency, $\Omega_{\rm K} = 1361$~Hz. The gravitational mass
  of the non-rotating star is $2.10 M_\odot$, which increases to $2.20
  M_\odot$ for rotation at $\Omega = \Omega_{\rm K}$. (Figure from Ref.~3.)
  \label{fig:profiles}}
\end{center}
\end{figure}

\begin{figure}[htb]
\begin{center}
  \includegraphics[scale=0.32]{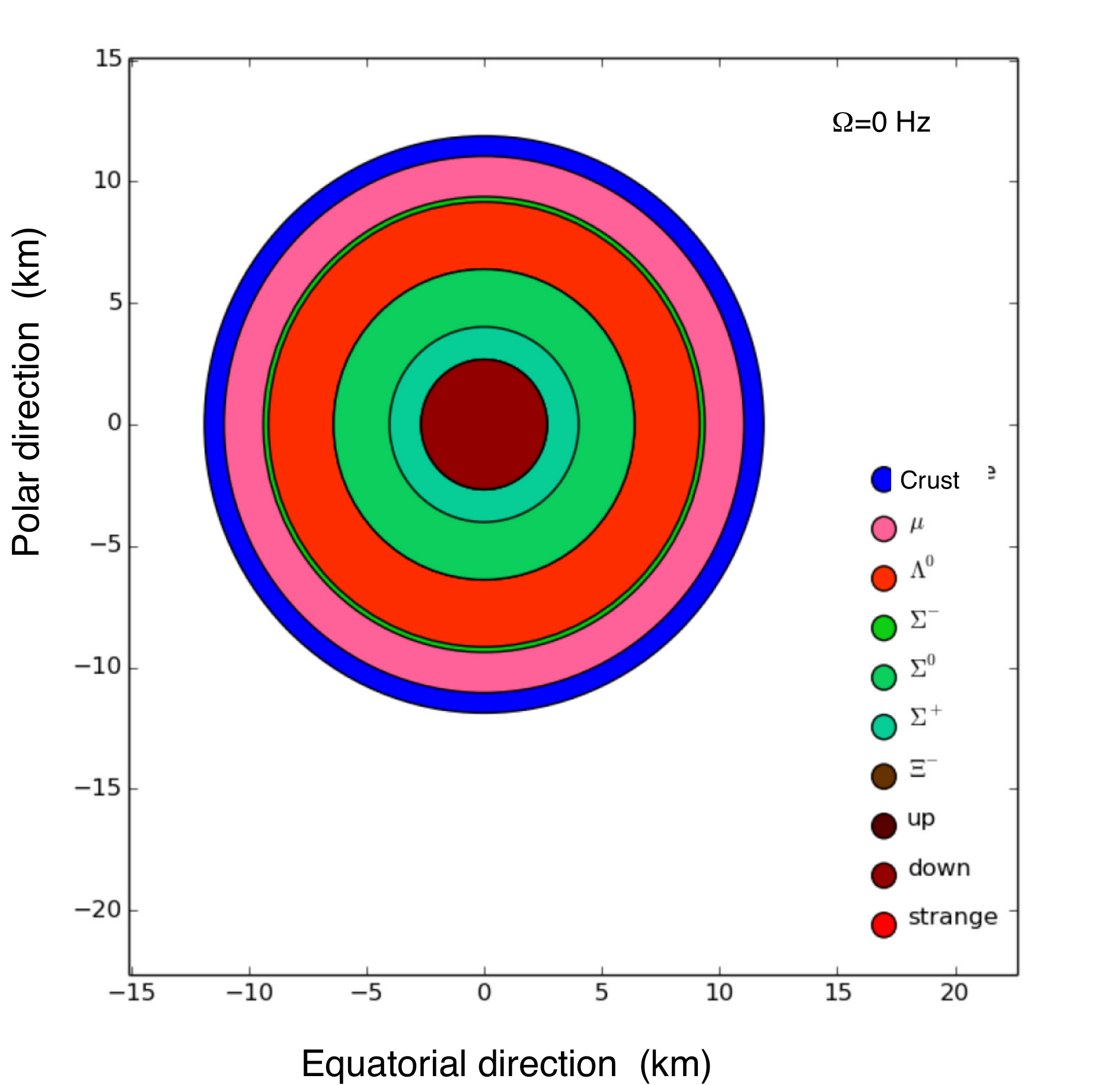}
      \includegraphics[scale=0.32]{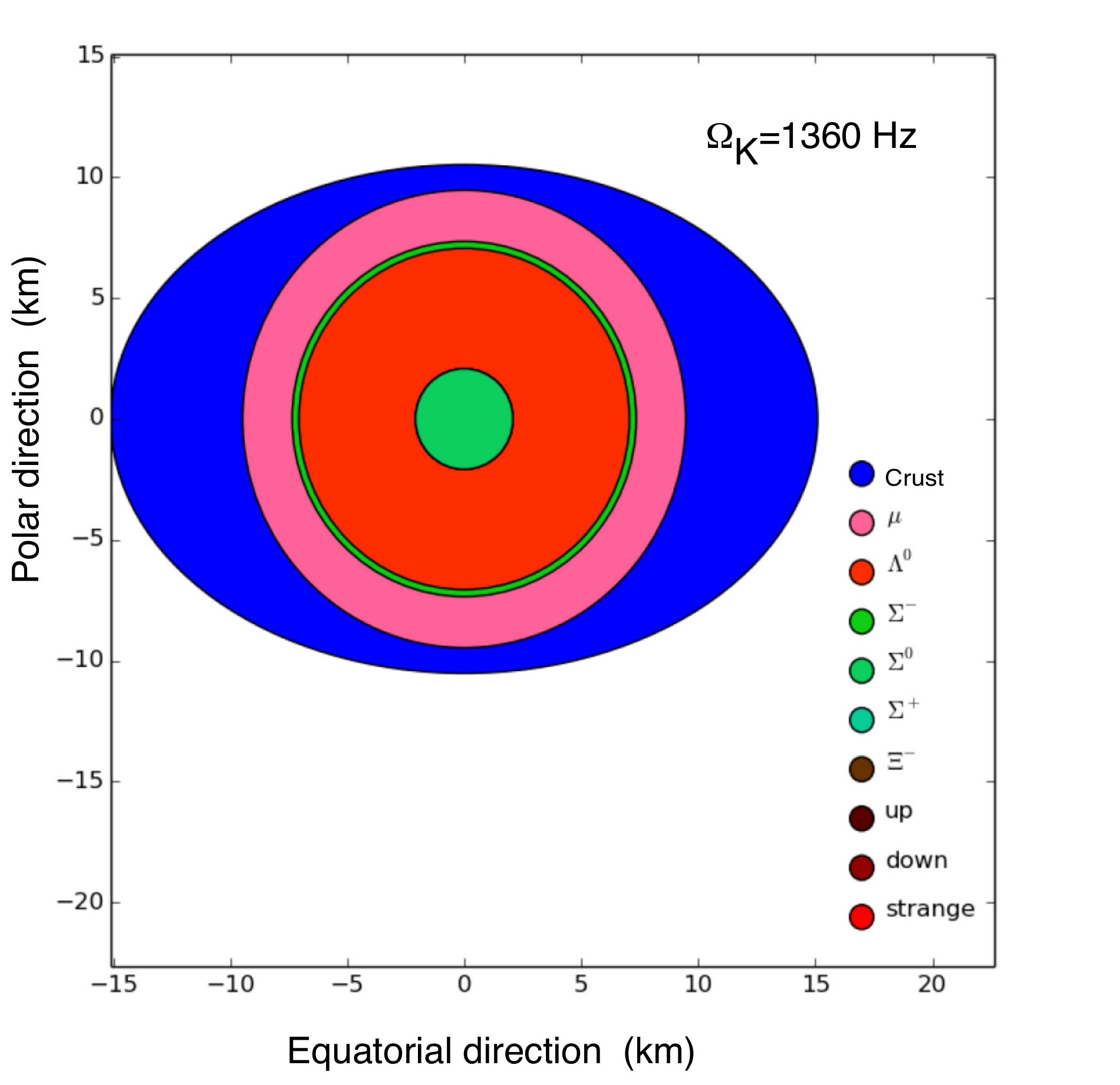}
\end{center}
\caption{Change of the composition of a $2 M_\odot$ neutron
  star caused by rotation, computed for the GM1 EoS.  The star on the
  left (right) hand-side is non-rotating (rotating at the Kepler
  frequency, $\Omega=\okgr$). The baryon number of both stars is the
  same ($\log_{10} A=57.51$). (Figure from Ref.~3.)}
  \label{fig:comp1}
\end{figure}

Figures \ref{fig:profiles} and \ref{fig:comp1} show the quark-hadron
compositions inside
of rotating neutron stars. These models contain extended regions of
matter consisting of a mixed phase of quarks and hadrons. This is a
consequence of the fact that the Gibbs condition has been used to
model the quark-hadron phase transition, in which case pressure varies
monotonically with the proportion of the phases in equilibrium. This
would not be the case if the Maxwell construction had been used to
model the phase equilibrium between quarks and hadrons. Currently, it
is an open issue which (if any of the two) descriptions characterizes
the phase transition in the cores of neutron stars properly. Among
other topics, this depends on the value of the surface tension of
quark matter.\cite{surfacetension:misc} As can be seen, the
quark-hadron mixed
phase as well as several different hyperon species are successively
spun out of the neutron star if the rotation rate increases toward the
Kepler frequency.  Non-rotating neutron stars posses the most complex
compositions, since they are the most dense members of the stellar
sequence. The compositions shown in Fig.\ \ref{fig:comp1} are
snapshots taken from movies  showing the entire rotational evolution of this neutron
star from zero frequency all the way to $\Omega_{\rm K}$. The movies are
publicly available.\cite{ref:movies} 

\begin{figure}[htb]
\begin{center}
\includegraphics[scale=0.45]{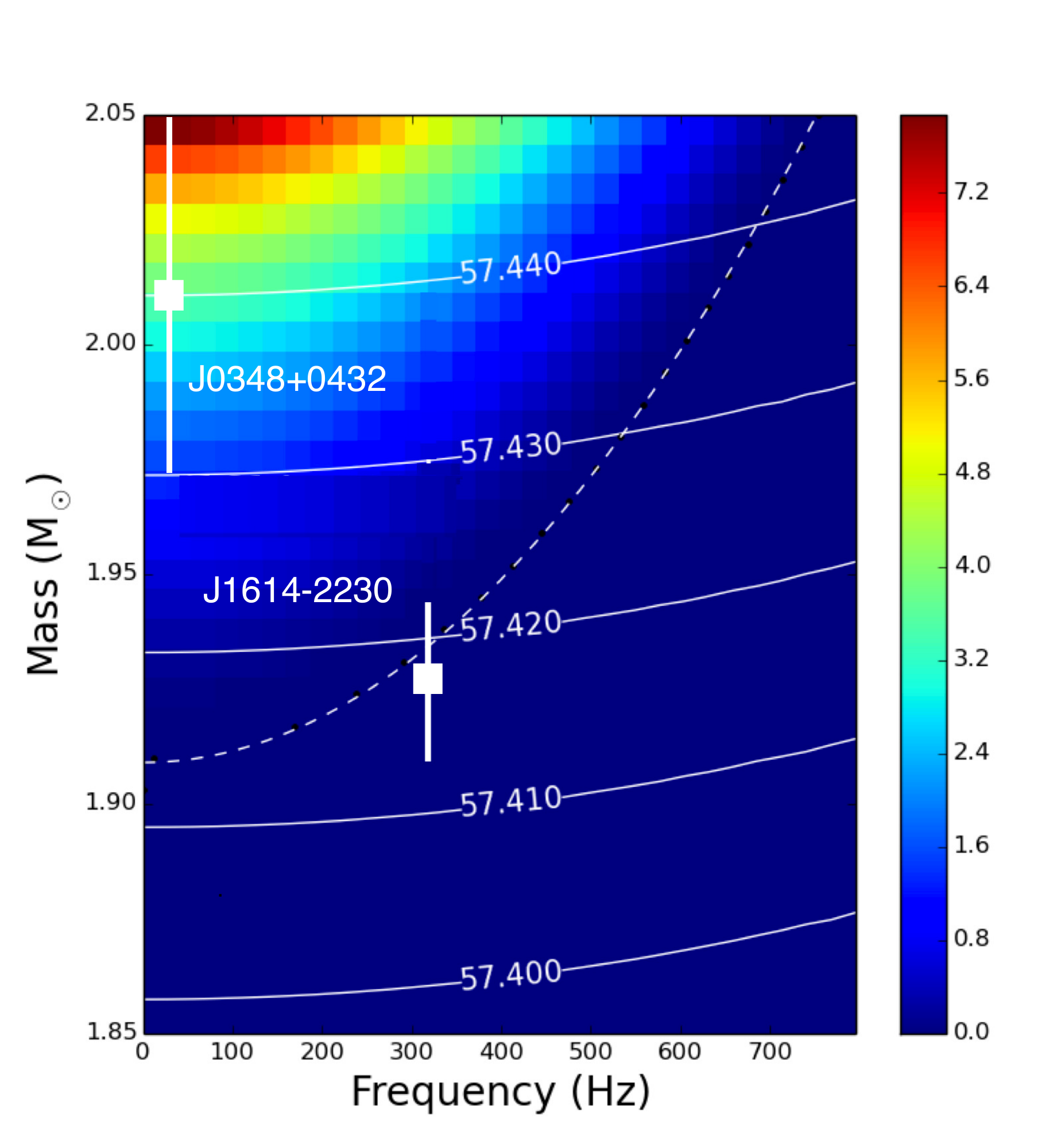}
\end{center}
\caption{Heat map showing the percent (column on the right) of total
  mass of a neutron star made up of deconfined quark matter, for the
  GM1 EoS. The white solid lines show the rotational evolution of
  neutron stars with constant baryon numbers $A$ (the reported figures
  being $\log_{10} A$). Also shown are the observed masses of pulsars
  J1614-2230 and J0348+0432 and the trend line (dashed white) fit,
  which separates confined from deconfined matter. (Figure from
  Ref.~3.)}
\label{GM1gv05QM}
\end{figure}

A heat map showing the quark-hadron content of rotating neutron stars
computed for the GM1 equation of state is shown in
Fig.\ \ref{GM1gv05QM}. As can be seen from this figure, up to around
8\% of the total gravitational mass of these neutron stars exists in
the form a mixed quark-hadron phase. Lines
of constant baryon number are also depicted in this figure as white
lines, labeled with the logarithm of the stars' baryon number. These
lines were included to give a sense of the path that a secluded
neutron star would be expected to take as it spins down.  By parsing
out the maximum frequency where deconfined quark matter is expected to
exist at the center of the neutron star for a given mass, it is
possible to get a curve through the gravitational mass--frequency
diagram for the threshold above which one can expect to find
deconfined quarks. These threshold frequencies for each mass were fit
to determine a (quadratic) function for the curve. The fit equation,
depicted as a dashed white line in Fig.\ \ref{GM1gv05QM}, was found to
have the form $M(\Omega) = a \Omega^2 + c$, where $M$ is the neutron
star's gravitational mass in solar masses, $\Omega$ its rotational
frequency, and $a$ and $c$ are parameters. The values for $a$ and $c$
can be found in Ref.\ \refcite{mellinger17:a}.

Figure \ref{GM1gv05QM} allows one to estimate the amount of
quark-hadron matter that may exist in the cores of neutron stars that
have both a measured frequency and mass, as illustrated for pulsars
PSR J1614--2230 ($M =1.928\pm 0.017 \, M_\odot$, rotational frequency
$f=318$~Hz)\cite{demorest10:a,fonseca16:a} and PSR J0348+0432 ($M=2.01
\pm 0.04\, M_\odot$, $f=26$~Hz)\cite{lynch13:a,antoniadis13:a} in
Figure \ref{GM1gv05QM}.  According to this calculation, up to around
7\% of the mass of PSR J0348+0432 could be in the mixed quark-hadron
phase, while the core of the more rapidly rotating pulsars PSR
J1614--2230 may be hovering right at the quark deconfinement density.

\section{Summary}\label{Sec:Discussion}

The true nature of the matter deep in the cores of rotating neutron
stars is still largely unknown despite several decades of intense
research on this topic.  In this short overview paper, we present the
results of recent\cite{mellinger17:a} neutron star calculations
performed for a non-local extension of the SU(3) Nambu--Jona-Lasinio
(NJL) model to investigate the possible existence of deconfined quarks
in the cores of neutron stars.  As shown in this paper, the type and
structure of the matter in the cores of rotating neutron stars depends
sensitively on the star's spin
frequency.\cite{weber99:book,weber05:a,stejner09:a} Exploring this
feature in more details opens up a new window on the type of matter
that exists in the central cores of neutron stars.  We find that,
depending on mass and rotational frequency, up to around 8\% of the
mass of massive neutron stars may be in the mixed quark-hadron phase,
if the quark-hadron phase transition is modeled as a Gibbs phase
transition. Examples of such massive neutron stars are pulsars PSR
J1614--2230 with a gravitational mass of $1.928\pm 0.017\,
M_{\odot}$\cite{demorest10:a,fonseca16:a} and PSR J0348+0432 with a
mass of $2.01 \pm 0.04 \, M_\odot$\cite{lynch13:a,antoniadis13:a}
(Fig.\ \ref{GM1gv05QM}).  Pure quark matter in the centers of neutron
stars is not obtained for any of the models for the nuclear equation
that we have studied.\cite{mellinger17:a} The gravitational mass at
which quark deconfinement sets in in rotating neutron stars varies
quadratically with spin frequency, which can be fitted by a simple
quadratic formula.  Owing to the unprecedented wealth of high-quality
data on pulsars provided by radio telescopes, X-ray satellites--and
soon the latest generation of gravitational-wave detectors--it seems
within reach to decode the inner workings of pulsars and, thus,
decipher the phase diagram of cold and ultra-dense hadronic matter
from astrophysics over the coming years.

\section*{Acknowledgments}
This work is supported through the U.S. National Science Foundation
under Grants PHY-1411708 and DUE-1259951.  M.G. Orsaria acknowledges
financial support from the American Physical Society's International
Research Travel Award Program.  G.A. Contrera and M.G. Orsaria
acknowledge financial support from CONICET and UNLP (Project
identification code 11/G140 and 11/X718), Argentina.  Computing
resources have been provided by the Computational Science Research Center
and the Department of Physics at San Diego State University.



\begin{thebibliography}{0}    

\bibitem{glen97:book} N. K. Glendenning, \textit{Compact Stars, Nuclear
    Physics, Particle Physics, and General Relativity}, 2nd ed.\
  (Springer-Verlag, New York, 2000).

\bibitem{weber99:book} F. Weber, \textit{Pulsars as Astrophysical
  Laboratories for Nuclear and Particle Physics}, High Energy Physics,
  Cosmology and Gravitation Series (IOP Publishing, Bristol, Great
  Britain, 1999).
  
\bibitem{mellinger17:a} R. D. Mellinger, F. Weber, W. Spinella,
  G. A. Contrera, and M. G. Orsaria, Universe {\bf 3}, 5 (2017).

\bibitem{contrera17:IWARA} G. A. Contrera, M. Orsaria,
  I. F. Ranea-Sandoval, and F. Weber, {\it Hybrid Stars in the
    Framework of different NJL Models.}, {\tt arXiv:1612.09485 [nucl-th]}.

\bibitem{friedman86:a} J. L. Friedman, J. R. Ipser, and L. Parker, 
  Astrophys.\ J.\ {\bf 304}, 115 (1986).

\bibitem{glen92:crust} N. K. Glendenning and F. Weber,
  Astrophys.\ J.\ {\bf 400}, 647 (1992).

\bibitem{glen92:limit} N. K. Glendenning, Phys.\ Rev.\ D {\bf 46},
  4161 (1992).
  
\bibitem{Hessels:2006ze} J. W. T. Hessels {\it et al.}, 
  Science {\bf 1901}, 311 (2006).
  
\bibitem{hessels06:a} J. W. T. Hessels, S. M. Ransom, I. H. Stairs, 
  P. C. C. Freire, V. M.  Kaspi, and F. Camilo, Science 
  {\bf 311}, 1901 (2006).

\bibitem{weber05:a}
F. Weber,  Prog. Part. Nucl. Phys. {\bf 54}, 193 (2005).
  
\bibitem{falcke14:a} H. Falcke and L. Rezzolla,
  A\&A {\bf 562}, A137 (2014).

\bibitem{surfacetension:misc} N.\ Yasutake {\it et al.}, Phys. Rev. C {\bf
  89} 065803 (2014); L.\ F.\ Palhares and E. S. Fraga, Phys. Rev. D
  {\bf 82}, 125018 (2010); M. B. Pinto, V. Koch, and J. Randrup,
  Phys. Rev. C {\bf 86}, 025203 (2012); B. W. Mintz, R. Stiele,
  R. O. Ramos, and J. Schaffner-Bielich, Phys. Rev. D {\bf 87}, 036004
  (2013).

\bibitem{ref:movies} {\tt http://www-rohan.sdsu.edu/~fweber/Mellinger/Compositions.html.}

\bibitem{demorest10:a} P. B. Demorest, T. Pennucci, S. M. Ranson, M. S. E.
  Roberts, and J. W. T.  Hessels, Nature {\bf 467},
  1081 (2010).

\bibitem{fonseca16:a} E. Fonseca {\it et al.}, Astrophys.\ J. {\bf
  832}, 2 (2016).

\bibitem{lynch13:a} R. S. Lynch {\it et al.}, Astrophys.\ J.\ {\bf 81},
  763 (2013),

\bibitem{antoniadis13:a} J. Antoniadis {\it et al.}, Science {\bf
   340}, 6131 (2013).

\bibitem{stejner09:a} M. Stejner, F. Weber, and J. Madsen, {\it
  Astrophys. J.} {\bf 694}, 1019 (2009).
  
\end{thebibliography}
\end{document}